\documentclass[aps,preprint]{revtex4}
\usepackage{amsfonts}
\usepackage{amsmath}
\usepackage{amssymb}
\usepackage{graphicx}

\input{tcilatex}

\begin{document}

\title{Exact soliton solutions and nonlinear modulation instability in
spinor Bose-Einstein condensates}
\author{Lu Li$^{1,2}$\thanks{%
e-mail llz@sxu.edu.cn}, Zaidong Li$^{1}$, Boris A. Malomed$^3$, Dumitru
Mihalache$^{4}$, and W. M. Liu$^{2}$}
\affiliation{$^{1}$College of Physics and Electronics Engineering, Shanxi University,
Taiyuan, 030006, China}
\affiliation{$^{2}$Joint Laboratory of Advanced Technology in Measurements, Beijing
National Laboratory for Condensed Matter Physics, Institute of Physics,
Chinese Academy of Sciences, Beijing 100080, China}
\affiliation{$^{3}$Department of Interdisciplinary Studies, Faculty of Engineering, Tel
Aviv University, Tel Aviv 69978, Israel}
\affiliation{$^{4}$National Institute of Physics and Nuclear Engineering, Institute of
Atomic Physics, Department of Theoretical Physics, P.O. Box MG-6, Bucharest,
Romania}

\begin{abstract}
We find one-, two-, and three-component solitons of the polar and
ferromagnetic (FM) types in the general (non-integrable) model of a spinor
(three-component) model of the Bose-Einstein condensate (BEC), based on a
system of three nonlinearly coupled Gross-Pitaevskii equations. The
stability of the solitons is studied by means of direct simulations, and, in
a part, analytically, using linearized equations for small perturbations.
Global stability of the solitons is considered by means of the energy
comparison. As a result, ground-state and metastable soliton states of the
FM and polar types are identified. For the special integrable version of the
model, we develop the Darboux transformation (DT). As an application of the
DT, analytical solutions are obtained that display full nonlinear evolution
of the modulational instability (MI) of a continuous-wave (CW) state seeded
by a small spatially periodic perturbation. Additionally, by dint of direct
simulations, we demonstrate that solitons of both the polar and FM types,
found in the integrable system, are structurally stable, i.e., they are
robust under random changes of the relevant nonlinear coefficient in time.
\end{abstract}

\maketitle

\section{Introduction}

The experimental observation of matter-wave solitons of the dark \cite%
{Burger}, bright \cite{bright}, and gap \cite{gap} types in Bose-Einstein
condensates (BECs) have attracted a great deal of attention to the dynamics
of nonlinear matter waves, including such aspects as the soliton propagation 
\cite{Busch}, vortex dynamics \cite{Rosenbusch}, interference patterns \cite%
{Liu}, domain walls in binary BECs \cite{DW}, and the modulational
instability (MI) \cite{Salasnich,Carr}.

One of major developments in BECs was the study of spinor condensates.
Spinor BECs feature an intrinsic three-component structure, due to the
distinction between different hyperfine spin states of the atoms. When
spinor BECs are trapped in the magnetic potential, the\ spin degree of
freedom is frozen. However, in the condensate held by an optical potential,
the spin is free, making it possible to observe a rich variety of phenomena,
such as spin domains \cite{Miesner} and textures \cite{Ohmi}. Recently,
properties of BECs with this degree of freedom were investigated in detail 
\cite{Stenger} - \cite{Ieda}, experimentally and theoretically. In
particular, the MI of a continuous-wave (CW) state with constant densities
in all the three components (sometimes, it is called ``driven
current\textquotedblright), was explored in Ref. \cite{MI}. An important
results was reported in Ref. \cite{Ieda}, which demonstrated that, under
special constraints imposed on parameters, the matrix nonlinear Schr\"{o}%
dinger (NLS) equation, which is a model of the one-dimensional (1D) spinor
BEC in the free space, may be integrable by means of the inverse scattering
transform. For that case, exact single-soliton solutions, as well as
solutions describing collisions between two solitons, were found.

In this work, we aim to consider several species of simple one-, two-, and
three-component soliton solutions, of both the ferromagnetic (FM) and polar
types, which can be found in an analytical form in the general
(non-integrable) version of the spinor model. Dynamical stability of the
solitons will be studied by means of analytical and numerical methods, and
global stability of different soliton types will be considered by comparison
of their energies, which will allow us to identify ground-state and
metastable soliton states. For the special integrable version of the model,
we will present a Darboux transformation (DT) and a family of exact
solutions generated by it from CW states. If the amplitude of the CW
background vanishes, these solutions go over into solitons found in Ref. 
\cite{Ieda}. The family also contains new spatially periodic time-dependent
solutions, that explicitly describe the full nonlinear development of the
MI, starting from a CW state seeded with an infinitely small spatially
periodic perturbation. In addition, we demonstrate structural stability of
the solitons of both polar and FM types in the integrable model, by showing
that they are robust under random changes of a nonlinear coefficient whose
value determines the integrability.

The paper is organized as follows. In Section 2, the model is formulated.
Various species of exact soliton solutions in the non-integrable model, and
their stability, are considered in Section 3. Section 4 introduces the DT,
presents solutions obtained by means of this technique, and also numerical
results demonstrating the structural stability of the solitons. The paper is
concluded by Section 5.

\section{The model}

We start with the consideration of an effectively one-dimensional BEC
trapped in a pencil-shaped region, which is elongated in $x$ and tightly
confined in the transverse directions $y,z$ (see, e.g., Ref. \cite{Carr}).
Therefore, atoms in the $F=1$ hyperfine state can be described by a 1D
vectorial wave function, $\mathbf{\Phi}(x,t)=[\Phi_{+1}(x,t),\Phi_{0}(x,t),%
\Phi _{-1}(x,t)]^{T}$, with the components corresponding to the three values
of the vertical spin projection, $m_{F}=+1,0,-1$. The wave functions obey a
system of coupled Gross-Pitaevskii (alias NLS) equations \cite{Ieda}, 
\begin{align}
i\hbar\partial_{t}\Phi_{\pm1} & =-\frac{\hbar^{2}}{2m}\partial_{x}^{2}\Phi_{%
\pm1}+(c_{0}+c_{2})(|\Phi_{\pm1}|^{2}+|\Phi_{0}|^{2})\Phi_{\pm 1}  \notag \\
& +(c_{0}-c_{2})|\Phi_{\mp1}|^{2}\Phi_{\pm1}+c_{2}\Phi_{\mp1}^{\ast}\Phi
_{0}^{2},  \notag \\
i\hbar\partial_{t}\Phi_{0} & =-\frac{\hbar^{2}}{2m}\partial_{x}^{2}\Phi
_{0}+(c_{0}+c_{2})(|\Phi_{+1}|^{2}+|\Phi_{-1}|^{2})\Phi_{0}  \notag \\
& +c_{0}|\Phi_{0}|^{2}\Phi_{0}+2c_{2}\Phi_{+1}\Phi_{-1}\Phi_{0}^{\ast},
\label{eqq}
\end{align}
where $c_{0}=(g_{0}+2g_{2})/3$ and $c_{2}=(g_{2}-g_{0})/3$ denote effective
constants of the mean-field (spin-independent) and spin-exchange
interaction, respectively \cite{Ho}. Here $g_{f}=4\hbar^{2}a_{f}/[ma_{%
\perp}^{2}(1-ca_{f}/a_{\perp})]$, with $f=0,2$, are effective 1D coupling
constants, $a_{f}$ is the $s$-wave scattering length in the channel with the
total hyperfine spin $f$, $a_{\perp}$ is the size of the transverse ground
state, $m$ is the atomic mass, and $c=-\zeta(1/2)\approx1.46$. Redefining
the wave function as $\mathbf{\Phi}\rightarrow(\phi_{+1},\sqrt{2}%
\phi_{0},\phi _{-1})^{T}$ and measuring time and length in units of $%
\hbar/|c_{0}|$ and $\sqrt{\hbar^{2}/2m|c_{0}|}$, respectively, we cast Eqs. (%
\ref{eqq}) in a normalized form 
\begin{align}
i\partial_{t}\phi_{\pm1} & =-\partial_{x}^{2}\phi_{\pm1}-(\nu+a)(|\phi
_{\pm1}|^{2}+2|\phi_{0}|^{2})\phi_{\pm1}  \notag \\
& -(\nu-a)|\phi_{\mp1}|^{2}\phi_{\pm1}-2a\phi_{\mp1}^{\ast}\phi_{0}^{2}, 
\notag \\
i\partial_{t}\phi_{0} & =-\partial_{x}^{2}\phi_{0}-2\nu|\phi_{0}|^{2}\phi_{0}
\notag \\
&
-(\nu+a)(|\phi_{+1}|^{2}+|\phi_{-1}|^{2})\phi_{0}-2a\phi_{+1}\phi_{-1}%
\phi_{0}^{\ast},  \label{eq}
\end{align}
where $\nu\equiv-\mathrm{\func{sgn}}\left( c_{0}\right) $, $%
a\equiv-c_{2}/|c_{0}|$.

These equations can be derived from the Hamiltonian%
\begin{align}
H & =\int_{-\infty}^{+\infty}dx\left\{ \left\vert \partial_{x}\phi
_{+1}\right\vert ^{2}+\left\vert \partial_{x}\phi_{-1}\right\vert ^{2}-\frac{%
1}{2}(\nu+a)\left( |\phi_{+1}|^{4}+|\phi_{-1}|^{4}\right)
-(\nu-a)|\phi_{+1}|^{2}|\phi_{-1}|^{2}\right.  \notag \\
& \left. +2\left[ \left\vert \partial_{x}\phi_{0}\right\vert
^{2}-\nu\left\vert \phi_{0}\right\vert ^{4}-(\nu+a)\left(
|\phi_{+1}|^{2}+|\phi_{-1}|^{2}\right) |\phi_{0}|^{2}-a\left(
\phi_{+1}^{\ast}\phi _{-1}^{\ast}\phi_{0}^{2}+\phi_{+1}\phi_{-1}\left(
\phi_{0}^{\ast}\right) ^{2}\right) \right] \right\} ,  \label{H}
\end{align}
which is a dynamical invariant of the model ($dH/dt=0$). In addition, Eqs. (%
\ref{eq}) conserve the momentum,%
\begin{equation*}
P=i\int_{-\infty}^{+\infty}\left(
\phi_{+1}^{\ast}\partial_{x}\phi_{+1}+\phi_{-1}^{\ast}\partial_{x}%
\phi_{-1}+2\phi_{0}^{\ast}\partial_{x}\phi _{0}\right) dx,
\end{equation*}
and the solution's norm, proportional to total number of atoms,%
\begin{equation}
N=\int_{-\infty}^{+\infty}\left[ \left\vert \phi_{+1}(x,t)\right\vert
^{2}+\left\vert \phi_{-1}(x,t)\right\vert ^{2}+2\left\vert
\phi_{0}(x,t)\right\vert ^{2}\right] dx.  \label{N}
\end{equation}

An obvious reduction of equations (\ref{eq}) can be obtained by setting $%
\phi_{0}\equiv0$. In this case, the model reduces to a system of two
equations which are tantamount to those describing light transmission in
bimodal nonlinear optical fibers, with the two modes representing either
different wavelengths or two orthogonal polarizations \cite{Agrawal}: 
\begin{equation}
i\partial_{t}\phi_{\pm1}=-\partial_{x}^{2}\phi_{\pm1}-(\nu+a)|\phi_{%
\pm1}|^{2}\phi_{\pm1} -(\nu-a)|\phi_{\mp1}|^{2}\phi_{\pm1}.  \label{fiber}
\end{equation}
In particular, the case of two linear polarizations in a birefringent fiber
corresponds to $(\nu-a)/(\nu+a)=2/3$, i.e., $a=\nu/5$, and two different
carrier waves or circular polarizations correspond to $(\nu-a)/(\nu+a)=2$,
i.e., $a=-\nu/3$. The two nonlinear terms in Eqs. (\ref{fiber}) account for,
respectively, the SPM and XPM (self- and cross-phase-modulation)
interactions of the two waves. The MI of CW (uniform) states in the system
of two XPM-coupled equations (\ref{fiber}) has been studied in detail \cite%
{Agrawal}. If the nonlinearity in Eqs. (\ref{fiber}) is self-defocusing,
i.e., $\nu=-1$ and $1\pm a>0$(in other words, $|a|~<1$), the single NLS\
equation would of course show no MI, but the XPM-coupled system of NLS
equations (\ref{fiber}) gives rise to MI, provided that the XPM interaction
is stronger than SPM, i.e., $0<a<1$.

\section{Exact single-, two-, and three-component soliton solutions}

The previous consideration of solitons in the model of the spinor BEC was
focused on the particular integrable case, $a=\nu=1$ \cite{Ieda}. As follows
from the above derivation, this special case is physically possible,
corresponding to $c_{2}=c_{0}$, or, equivalently, $2g_{0}=-g_{2}>0$. The
latter condition can be satisfied by imposing a condition on the scattering
lengths,%
\begin{equation}
a_{\perp}=3ca_{0}a_{2}/(2a_{0}+a_{2}),  \label{scattlength}
\end{equation}
provided that $a_{0}a_{2}(a_{2}-a_{0})>0$ holds.

However, it is also possible to find simple exact soliton solutions of both
polar and FM types in the general case, $a\neq\nu$. Below, we present
several soliton species for this case, and study their stability.

\subsection{Single-component ferromagnetic soliton}

First, a single-component FM soliton is given by a straightforward solution
(provided that $a+\nu>0$),%
\begin{equation}
\phi_{-1}=\phi_{0}=0,\text{ }\phi_{+1}=\sqrt{\frac{-2\mu}{\nu+a}}\mathrm{%
\func{sech}}\left( \sqrt{-\mu}x\right) e^{-i\mu t},  \label{ferro}
\end{equation}
where the negative chemical potential $\mu$ is the intrinsic parameter of
the soliton family. While the expression (\ref{ferro}) corresponds to a
zero-velocity soliton, moving ones can be generated from it in an obvious
way by means of the Galilean transformation. Note that the condition $%
a+\nu>0 $ implies $a>1$ or $a>-1$ in the cases of the, respectively,
repulsive ($\nu =-1$) and attractive ($\nu=+1$) spin-independent
interaction. The norm (\ref{N}) and energy (\ref{H}) of this soliton are%
\begin{equation}
N=\frac{4\sqrt{-\mu}}{\nu+a},\text{ }H=-\frac{\left( \nu+a\right) ^{2}}{48}%
N^{3}.  \label{NEferro}
\end{equation}

The soliton (\ref{ferro}) is \emph{stable} against infinitesimal
perturbations, as the linearization of Eqs. (\ref{eq}) about this solution
demonstrates that all the three equations for small perturbations are \emph{%
decoupled}. Then, as the standard soliton solution of the single NLS
equation is always stable, solution (\ref{ferro}) cannot be unstable against
small perturbations of $\phi _{+1}$. Further, the decoupled linearized
equations for small perturbations $\phi _{-1}$ and $\phi _{0}$ of the other
fields are%
\begin{align}
i\partial _{t}\phi _{-1}& =-\partial _{x}^{2}\phi _{-1}-(\nu -a)|\phi
_{+1}|^{2}\phi _{-1}, \\
i\partial _{t}\phi _{0}& =-\partial _{x}^{2}\phi _{0}-(\nu +a)|\phi
_{+1}|^{2}\phi _{0},
\end{align}%
and it is well known that such linear equations, with $|\phi _{+1}|^{2}$
corresponding to the unperturbed soliton (\ref{ferro}), give rise to no
instabilities either.

\subsection{Single-component polar soliton}

The simplest polar soliton, that has only the $\phi _{0}$ component, can be
found for $\nu =+1$ (note that the solution does not depend on the other
nonlinear coefficient $a$):%
\begin{equation}
\phi _{0}=\sqrt{-\mu }\mathrm{\func{sech}}\left( \sqrt{-\mu }x\right)
e^{-i\mu t},\text{ }\phi _{\pm 1}=0.~  \label{polar1}
\end{equation}%
The stability problem for this soliton is more involved, as the unperturbed
field $\phi _{0}$ gives rise to a \emph{coupled} system of linearized
equations for infinitesimal perturbations $\phi _{\pm 1}$ of the other
fields:%
\begin{equation}
\left( i\partial _{t}+\mu \right) \chi _{\pm 1}=-\partial _{x}^{2}\chi _{\pm
1}+2\mu \mathrm{\func{sech}}^{2}\left( \sqrt{-\mu }x\right) \left[ (1+a)\chi
_{\pm 1}+a\chi _{\mp 1}^{\ast }\right] ,  \label{pert-polar1}
\end{equation}%
where we have substituted $\nu =+1$ and redefined the perturbation, 
\begin{equation}
\phi _{\pm 1}\equiv \exp \left( -i\mu t\right) \chi _{\pm 1}.  \label{redef}
\end{equation}%
The last term in Eqs. (\ref{pert-polar1}) gives rise to a \textit{parametric
gain}, which may be a well-known source of instability (see, e.g., Ref. \cite%
{Barash}). This instability can be easily understood in qualitative terms if
one replaces Eqs. (\ref{pert-polar1}) by simplified equations, that
disregard the $x$-dependence, and replace the waveform $\mathrm{\func{sech}}%
^{2}\left( \sqrt{-\mu }x\right) $ by its value at the soliton's center, $x=0$%
:%
\begin{equation}
i\frac{d\chi _{\pm 1}}{dt}=\mu \left[ \left( 1+2a\right) \chi _{\pm
1}+2a\chi _{\mp 1}^{\ast }\right] .  \label{simplified}
\end{equation}%
An elementary consideration shows that the zero solution of linear equations
(\ref{simplified}) is double-unstable (i.e., through a double eigenvalue) in
the region of $a<-1/4$.

We checked the stability of the single-component polar soliton (\ref{polar1}%
) by means of direct simulations of Eqs. (\ref{eq}), adding a small random
perturbation in the components $\phi _{+1}$ and $\phi _{-1}$, the value of
the perturbation being distributed uniformly between $0$ and $0.03$ (the
same random perturbation will be used in simulations of the stability of
other solitons, see below). The result is that the soliton (\ref{polar1}) is
always unstable indeed, as an example in Fig. 1. In particular, panel (d) in
this figure displays the time evolution of 
\begin{equation}
N_{0}=\int_{-\infty }^{+\infty }\left\vert \phi _{0}(x,t)\right\vert
^{2}dx,N_{\pm 1}=\int_{-\infty }^{+\infty }\left\vert \phi _{\pm
}(x,t)\right\vert ^{2}dx.  \label{NN}
\end{equation}

\subsection{Two-component polar soliton}

In the same case as considered above, $\nu =+1$, a two-component polar
soliton can be easily found too:%
\begin{equation}
\phi _{0}=0,\text{ }\phi _{+1}=\pm \phi _{-1}=\sqrt{-\mu }\mathrm{\func{sech}%
}\left( \sqrt{-\mu }x\right) e^{-i\mu t}  \label{polar2}
\end{equation}%
(note that it again does depend on the value of the nonlinearity coefficient 
$a$). In this case again, instability is possible due to the parametric gain
in linearized equations for small perturbations. Indeed, the corresponding
equation for the perturbation in the $\phi _{0}$ component decouples from
the other equations and takes the form%
\begin{equation}
\left( i\partial _{t}+\mu \right) \chi _{0}=-\partial _{x}^{2}\chi _{0}+2\mu 
\text{ }\mathrm{\func{sech}}^{2}\left( \sqrt{-\mu }x\right) \left[ (1+a)\chi
_{0}\pm a\chi _{0}^{\ast }\right] ,  \label{pert-polar2}
\end{equation}%
where the perturbation was redefined the same way as in Eq. (\ref{redef}), $%
\phi _{0}\equiv \exp \left( -i\mu t\right) \chi _{0}$. Similar to the case
of Eq. (\ref{pert-polar1}), the origin of the instability may be understood,
replacing Eq. (\ref{pert-polar2}) by its simplified version that disregards
the $x$-dependence and substitutes the waveform $\mathrm{\func{sech}}%
^{2}\left( \sqrt{-\mu }x\right) $ by its value at the central point of the
soliton, $x=0$:%
\begin{equation}
i\frac{d\chi _{0}}{dt}=\mu \left[ \left( 1+2a\right) \chi _{0}\pm 2a\chi
_{0}^{\ast }\right] ,  \label{simplified2}
\end{equation}%
cf. Eqs. (\ref{simplified}). An elementary consideration demonstrates that
the zero solution of Eq. (\ref{simplified2}) (for either sign $\pm $) is
unstable in exactly the same region as in the case of Eqs. (\ref{simplified}%
), $a<-1/4$. Note that the simplified equation (\ref{simplified2})
illustrates only the qualitative mechanism of the parametric instability,
while the actual stability border may be different from $a=-1/4.$ Indeed,
the stability of the soliton (\ref{polar2}) was tested by direct simulations
of Eqs. (\ref{eq}) with a small uniformly distributed random perturbation
added to the $\phi _{0}$ component. The result shows that the soliton is
indeed unstable in the region of $|a|~\geq 1$, as shown in Fig. 2, and it is
stable if $\left\vert a\right\vert <1$ (not shown here).\ \ 

The change of the stability of this soliton at $a=+1$ can be explained
analytically. Indeed, splitting the perturbation into real and imaginary
parts, $\chi _{0}(x,t)\equiv \chi _{1}(x,t)+i\chi _{2}(x,t)$, and looking
for a perturbation eigenmode as $\chi _{1,2}(x,t)=e^{\sigma t}U_{1,2}(x)$
with the instability growth rate (eigenvalue) $\sigma $, one arrives at a
system of real equations [take plus sign in (\ref{pert-polar2})],%
\begin{align}
-\sigma U_{2}& =-U_{1}^{\prime \prime }-\mu U_{1}+2\mu (1+2a)\mathrm{\func{%
sech}}^{2}\left( \sqrt{-\mu }x\right) U_{1},  \notag \\
\sigma U_{1}& =-U_{2}^{\prime \prime }-\mu U_{2}+2\mu ~\mathrm{\func{sech}}%
^{2}\left( \sqrt{-\mu }x\right) U_{2},  \label{sigma}
\end{align}%
with prime standing for $d/dx$. The simplest possibility for the onset of
instability of the soliton is the passage of $\sigma $ from negative values,
corresponding to neutral stability, to unstable positive values. At the
critical (zero-crossing) point, $\sigma =0$, equations (\ref{sigma})
decouple, and each one becomes explicitly solvable, as is commonly known
from quantum mechanics. The exact solution of the decoupled equations shows
that $\sigma $ crosses zero at points $a=a_{n}\equiv n(n+3)/4$, $n=0,1,2,...$%
. The vanishing of $\sigma $ at the point $a_{0}=0$ corresponds not to
destabilization of the soliton, but just to the fact that Eqs. (\ref{sigma})
become symmetric at this point, while the zero crossings at other points
indeed imply stability changes. In particular, the critical point $a_{1}=1$
explains the destabilization of soliton (\ref{polar2}) at $a=1$, as observed
in the simulations, the corresponding eigenfunctions being $U_{1}=\sinh
\left( \sqrt{-\mu }x\right) \cdot \mathrm{sech}^{2}\left( \sqrt{-\mu }%
x\right) $, $U_{2}=\mathrm{sech}\left( \sqrt{-\mu }x\right) $. Critical
points corresponding to $n>1$ imply, most probably, additional
destabilizations (the emergence of new unstable eigenmodes) of the already
unstable soliton. On the other hand, the destabilization of the soliton at $%
a=-1$, also observed in the simulations, may be explained by a more involved
bifurcation, when a pair of the eigenvalues $\sigma $ change from purely
imaginary to complex ones, developing an unstable (positive) real part (a
generic route to the instability of the latter type is accounted for by the 
\textit{Hamiltonian Hopf bifurcation}, which involves a collision between
two pairs of imaginary eigenvalues, and their transformation into an
unstable quartet \cite{HH}).

\subsection{Three-component polar solitons}

In the case of $\nu =+1$, three-component solitons of the polar type can be
found too. One of them is%
\begin{eqnarray}
\phi _{0} &=&\sqrt{1-\epsilon ^{2}}\sqrt{-\mu }~\mathrm{\func{sech}}\left( 
\sqrt{-\mu }x\right) e^{-i\mu t},\text{ }  \notag \\
\phi _{+1} &=&-\phi _{-1}=\pm \epsilon \sqrt{-\mu }~\mathrm{\func{sech}}%
\left( \sqrt{-\mu }x\right) e^{-i\mu t},  \label{polar31}
\end{eqnarray}%
where $\epsilon $ is an arbitrary parameter taking values $-1<\epsilon <+1$
(the presence of this parameter resembles the feature typical to solitons in
the Manakov's system \cite{Manakov,Agrawal}), and, as well as the one- and
two-component polar solitons displayed above, the solution does not
explicitly depend on the parameter $a$. We stress that the phase difference
of $\pi $ between the $\phi _{+1}$ and $\phi _{-1}$ components is a
necessary ingredient of the solution.

There is another three-component polar solution similar to the above one
(i.e., containing the arbitrary parameter $\epsilon $, and independent of $a$%
), but with equal phases of the $\phi _{\pm 1}$ components and a phase shift
of $\pi /2$ in the $\phi _{0}$ component. This solution is%
\begin{eqnarray}
\phi _{0} &=&i\sqrt{1-\epsilon ^{2}}\sqrt{-\mu }~\mathrm{\func{sech}}\left( 
\sqrt{-\mu }x\right) e^{-i\mu t},\text{ }  \notag \\
\phi _{+1} &=&\phi _{-1}=\pm \epsilon \sqrt{-\mu }~\mathrm{\func{sech}}%
\left( \sqrt{-\mu }x\right) e^{-i\mu t},  \label{polar32}
\end{eqnarray}%
where the sign $\pm $ is the same for both components.

In addition, there is a species of three-component polar solitons that
explicitly depend on $a$:%
\begin{eqnarray}
\phi _{0} &=&\left( \mu _{+1}\mu _{-1}\right) ^{1/4}~\mathrm{\func{sech}}%
\left( \sqrt{-\mu }x\right) e^{-i\mu t},  \notag \\
\phi _{\pm 1} &=&\sqrt{-\mu _{\pm 1}}~\mathrm{\func{sech}}\left( \sqrt{-\mu }%
x\right) e^{-i\mu t},  \label{polar3a}
\end{eqnarray}%
where $\mu _{\pm 1}$ are two arbitrary negative parameters, and the chemical
potential is $\mu =-\left( \nu +a\right) \left( \sqrt{-\mu _{+1}}+\sqrt{-\mu
_{-1}}\right) ^{2}/2$, which implies that $\nu +a>0$ (note that this
solution admits $v=-1$, i.e., repulsive spin-independent interaction). Each
species of the three-component polar soliton depends on two arbitrary
parameters: either $\mu $ and $\epsilon $ [solutions (\ref{polar31}) and (%
\ref{polar32})], or $\mu _{-1}$ and $\mu _{+1}$ [solution (\ref{polar3a})].

The stability of all the species of these solitons was tested in direct
simulations. In the case of solutions (\ref{polar31}) and (\ref{polar32}),
it was enough to seed a small random perturbation only in the $\phi _{0}$
component, to observe that both these types are unstable, as shown in Fig. 3
and 4.

On the contrary, the three-component polar soliton (\ref{polar3a}) is \emph{%
completely stable}. This point was checked in detail, by seeding small
random perturbations in all the components. The result, illustrated by Fig.
5, is that the perturbation induces only small oscillations of amplitudes of
each component of the soliton, and remain robust as long as the simulations
were run.

\subsection{Global stability of the solitons}

It follows from the above results that the model demonstrates remarkable
multistability, as the FM soliton (\ref{ferro}), the two-component polar
solitons (\ref{polar2}) (in the regions of $-1<a<+1$), and the
three-component soliton (\ref{polar3a}) may all be stable in one and the
same parameter region. Then, to identify which solitons are
\textquotedblleft more stable\textquotedblright\ and \textquotedblleft less
stable\textquotedblright , one may fix the norm (\ref{N}) and compare
respective values of the Hamiltonian (\ref{H}) for these solutions, as the
ground-state solution should correspond to a minimum of $H$ at given $N$.

The substitution of the solutions in Eq. (\ref{H}) reveals another
remarkable fact: all the solutions which do not explicitly depend on $a$,
i.e., the one-, two-, and three-component polar solitons (\ref{polar1}), (%
\ref{polar2}), and (\ref{polar31}), (\ref{polar32}), produce identical
relations between the chemical potential ($\mu$), norm, and Hamiltonian:%
\begin{equation}
N=4\sqrt{-\mu},\text{ }H=-\frac{1}{48}N^{3}.  \label{NEpolar}
\end{equation}
As mentioned above, these relations are different for the single-component
FM soliton (\ref{ferro}), see Eqs. (\ref{NEferro}). Finally, for the stable
three-component polar soliton (\ref{polar3a}), the relations between $\mu$, $%
N$, and $H$ take exactly the same form as Eqs. (\ref{NEferro}) for the FM
soliton.

The comparison of the expressions (\ref{NEpolar}) and (\ref{NEferro}) leads
to a simple conclusion: the single-component FM soliton (\ref{ferro}) and
the stable three-component polar soliton (\ref{polar3a}) \emph{simultaneously%
} provide for the minimum of energy in the case of $\nu=+1$ and $a>0$, i.e.,
when both the spin-independent and spin-exchange interactions between atoms
are attractive (recall a necessary condition for the existence of both these
types of the solitons is $\nu+a>0$). It is possible to state that, in these
cases, each of these two species, (\ref{ferro}) and (\ref{polar3a}), plays
the role of the ground state in its own class of the solitons (ferromagnetic
and polar ones, respectively). Simultaneously, the two-component soliton (%
\ref{polar2}) is also stable in the region of $0<a<1$, as shown above, but
it corresponds to higher energy, hence it represents a metastable state in
this region.

The FM soliton (\ref{ferro}) and three-component polar soliton (\ref{polar3a}%
) also exist in the case of $\nu=-1$ and $a>1$, when $\nu+a$ is positive. In
this case too, these two soliton species provide for the energy minimum,
simply because the other solitons, (\ref{polar1}), (\ref{polar2}), (\ref%
{polar31}), and (\ref{polar32}) do not exist at $\nu=-1$.

For $\nu=+1$ and $-1<a<0$ (attractive spin-independent and repulsive
spin-exchange interactions), the Hamiltonian value (\ref{NEpolar}) is
smaller than the competing one (\ref{NEferro}). Actually, this means that
the two-component polar soliton (\ref{polar2}) plays the role of the ground
state in this case, as only it, among all the polar solitons whose
Hamiltonian is given by the expression (\ref{NEpolar}), is dynamically
stable (for $-1<a<+1$, see above). As the FM soliton (\ref{ferro}) and
three-component polar soliton (\ref{polar3a}) also exist and are stable in
this region, but correspond to greater energy, they are metastable states
here. Finally, for $\nu=+1$ and $a<-1$, there are no stable solitons (among
the ones considered above).

The dependence $N(\mu )$ for each solution family provides additional
concerning the soliton stability. Indeed, the known Vakhitov-Kolokolov (VK)
criterion \cite{VK} states that a necessary stability condition for the
soliton family is $dM/d\mu <0$. It guarantees that the soliton cannot be
unstable against perturbations with real eigenvalues, but does not say
anything about oscillatory perturbation modes appertaining to complex
eigenvalues. Obviously, both relations, given by Eqs. (\ref{NEferro}) and (%
\ref{NEpolar}), satisfy the VK condition, i.e., any unstable soliton may be
unstable only against perturbations that grow in time with oscillations.
Indeed, numerical results which illustrate the instability of those solitons
which are not stable, see Figs. 1-4, clearly suggest that the instability,
if any, is oscillatory.

\subsection{Finite-background solitons}

In special cases, it is possible to find exact solutions for solitons
sitting on a nonzero background. Namely, for $\nu=1$ and $a=-1/2$, one can
find a two-component polar soliton with a continuous-wave (CW) background
attached to it, in the following form: $\phi_{0}=0$, and 
\begin{align}
\phi_{+1} & =e^{-i\mu t}\sqrt{-\mu}\left[ \frac{1}{\sqrt{2}}\pm \mathrm{%
\func{sech}}\left( \sqrt{-\mu}x\right) \right] ,  \notag \\
\phi_{-1} & =e^{-i\mu t}\sqrt{-\mu}\left[ \frac{1}{\sqrt{2}}\mp \mathrm{%
\func{sech}}\left( \sqrt{-\mu}x\right) \right].  \label{polarb2}
\end{align}
For $\nu=a=1$, a three-component polar solution with the background can be
found too: 
\begin{align}
\phi_{+1} & =\phi_{-1}=\frac{1}{2}\sqrt{-\mu}e^{-i\mu t}\left[ \frac {1}{%
\sqrt{2}}\pm\mathrm{\func{sech}}\left( \sqrt{-\mu}x\right) \right] ,  \notag
\\
\phi_{0} & =\frac{1}{2}\sqrt{-\mu}e^{-i\mu t}\left[ \frac{1}{\sqrt{2}}\mp%
\mathrm{\func{sech}}\left( \sqrt{-\mu}x\right) \right]  \label{polarb3}
\end{align}
(in the latter case, the availability of the exact solution is not
surprising, as the case of $\nu=a=1$ is the exactly integrable one \cite%
{Ieda}).

The stability of solutions (\ref{polarb2}) and (\ref{polarb3}) was tested,
simultaneously perturbing the $\phi _{0}$ and $\phi _{\pm 1}$ components by
small uniformly distributed random perturbations. Figures 6 and 7 show that
both solutions are unstable, although the character of the instability is
different: in the former case, the soliton itself (its core) seems stable,
while the background appears to be modulationally unstable. In the latter
case (Fig. 7), the background is modulationally stable, but the core of the
soliton (\ref{polarb3}) clearly features an oscillatory instability.

The modulational instability (MI)\ of the background in solution (\ref%
{polarb2}) is obvious, as, even without exciting the $\phi _{0}$ field,
i.e., within the framework of the reduced equations (\ref{fiber}), with $\nu
+a=1/2$ and $\nu -a=3/2$, any CW solution is obviously subject to MI. As
concerns the excitation of the $\phi _{0}$ field, an elementary
consideration shows that the CW part of solution (\ref{polarb2}) exactly
corresponds to the threshold of the instability possible through the action
of the parametric-drive term, $2a\phi _{+1}\phi _{-1}\phi _{0}^{\ast }$, in
the last equation (\ref{eq}). Actually, Fig. 6 demonstrate that this
parametric instability sets in, which may be explained by conjecturing that
the initial development of the above-mentioned MI that does not include the $%
\phi _{0}$ field drives the perturbed system across the threshold of the
parametric instability.

\section{The Darboux transformation and nonlinear development of the
modulational instability}

In this section, we focus on the integrable case, with $\nu =a=1$, which
corresponds to the attractive interactions. As explained above, the spinor
BEC obeys this condition if a special (but physically possible) constraint (%
\ref{scattlength}) is imposed on the scattering lengths which determine
collisions between atoms. Then, Eqs. (\ref{eq}) can be rewritten as a $%
2\times 2$ matrix NLS equation \cite{Ieda,Tsuchida}, 
\begin{equation}
i\partial _{t}\mathbf{Q}+\partial _{x}^{2}\mathbf{Q}+2\mathbf{QQ}^{\dagger }%
\mathbf{Q}=0,~\mathbf{Q}\equiv \left( 
\begin{array}{cc}
\phi _{+1} & \phi _{0} \\ 
\phi _{0} & \phi _{-1}%
\end{array}%
\right) .  \label{eq1}
\end{equation}%
Equation (\ref{eq1}) is a completely integrable system, to which the inverse
scattering transform applies \cite{Tsuchida} [in particular, with regard to
the fact that $\nu -a=0$ in the present case, Eqs. (\ref{fiber}) show that
system (\ref{eq1}) splits into two uncoupled NLS equations for solutions
with $\phi _{0}=0$]. We here aim to develop the Darboux transformation (DT)
for Eqs. (\ref{eq1}), based on the respective Lax pair.

The Lax-pair representation of Eqs. (\ref{eq1}) is \cite{Tsuchida} 
\begin{equation}
\mathbf{\Psi}_{x}=\mathbf{U\Psi},~\mathbf{\Psi}_{t}=\mathbf{V\Psi},
\label{lax}
\end{equation}
where $\mathbf{U}=\lambda\mathbf{J}+\mathbf{P}$, $\mathbf{V}=i2\lambda ^{2}%
\mathbf{J}+i2\lambda\mathbf{P}+i\mathbf{W}$, with 
\begin{align}
\mathbf{J} & =\left( 
\begin{array}{cc}
\mathbf{I} & \mathbf{O} \\ 
\mathbf{O} & -\mathbf{I}%
\end{array}
\right) ,\mathbf{P}=\left( 
\begin{array}{cc}
\mathbf{O} & \mathbf{Q} \\ 
-\mathbf{Q}^{\dagger} & \mathbf{O}%
\end{array}
\right) , \\
\mathbf{W} & =\left( 
\begin{array}{cc}
\mathbf{QQ}^{\dagger} & \mathbf{Q}_{x} \\ 
\mathbf{Q}_{x}^{\dagger} & -\mathbf{Q}^{\dagger}\mathbf{Q}%
\end{array}
\right) .
\end{align}
Here $\mathbf{I}$ denotes the $2\times2$ unit matrix, $\mathbf{O}$ is the $%
2\times2$ zero matrix, $\mathbf{\Psi}=\left( 
\begin{array}{c}
\mathbf{\Psi}_{1} \\ 
\mathbf{\Psi}_{2}%
\end{array}
\right) $ is the matrix eigenfunction corresponding to $\lambda$, $\mathbf{%
\Psi}_{1}$ and $\mathbf{\Psi}_{2}$ being $2\times2$ matrices, and $\lambda$
is the spectral parameter. Accordingly, Eq. (\ref{eq1}) can be recovered
from the compatibility condition of the linear system (\ref{lax}), $\mathbf{U%
}_{t}-\mathbf{V}_{x}+\mathbf{UV}-\mathbf{VU}=0$.

Next, based on the Lax pair (\ref{lax}), and introducing a transformation in
the form 
\begin{equation}
\widetilde{\mathbf{\Psi}}=(\lambda-\mathbf{S})\mathbf{\Psi},~\mathbf{S}=%
\mathbf{D\Lambda D}^{-1},~\mathbf{\Lambda}=\left( 
\begin{array}{cc}
\lambda_{1}\mathbf{I} & \mathbf{O} \\ 
\mathbf{O} & \lambda_{2}\mathbf{I}%
\end{array}
\right) ,
\end{equation}
where $\mathbf{D}$ is a nonsingular matrix that satisfies $\mathbf{D}_{x}=%
\mathbf{JD\Lambda}+\mathbf{PD}$, and letting 
\begin{equation}
\widetilde{\mathbf{\Psi}}_{x}=\widetilde{\mathbf{U}}\widetilde{\mathbf{\Psi}}%
,~\widetilde{\mathbf{U}} =\lambda\mathbf{J}+\mathbf{P}_{1},~\mathbf{P}%
_{1}\equiv\left( 
\begin{array}{cc}
\mathbf{O} & \mathbf{Q}_{1} \\ 
-\mathbf{Q}_{1}^{\dagger} & \mathbf{O}%
\end{array}
\right) ,
\end{equation}
one obtains%
\begin{equation}
\mathbf{P}_{1}=\mathbf{P}+\mathbf{JS}-\mathbf{SJ}.  \label{db1}
\end{equation}
Also, one can verify the following \textit{involution} property of the above
linear equations: if $\mathbf{\Psi}=\left( 
\begin{array}{c}
\mathbf{\Psi}_{1} \\ 
\mathbf{\Psi}_{2}%
\end{array}
\right) $ is an eigenfunction corresponding to $\lambda$, where $\mathbf{\Psi%
}_{j}$ are $2\times2$ matrices, then $\left( 
\begin{array}{c}
-\mathbf{\Psi}_{2}^{\ast} \\ 
\mathbf{\Psi}_{1}^{\ast}%
\end{array}
\right) $ is an eigenfunction corresponding to $-\lambda^{\ast}$, where we
have used the fact that $\mathbf{Q}^{\ast}=\mathbf{Q}^{\dag}$ [the matrix $%
\mathbf{Q}$ is symmetric, as per its definition (\ref{eq1})]. Thus we can
take $\mathbf{D}$ in the form 
\begin{equation*}
\mathbf{D}=\left( 
\begin{array}{cc}
\mathbf{\Psi}_{1} & -\mathbf{\Psi}_{2}^{\ast} \\ 
\mathbf{\Psi}_{2} & \mathbf{\Psi}_{1}^{\ast}%
\end{array}
\right) ,~\mathbf{\Lambda}=\left( 
\begin{array}{cc}
\lambda\mathbf{I} & \mathbf{O} \\ 
\mathbf{O} & -\lambda^{\ast}\mathbf{I}%
\end{array}
\right) ,
\end{equation*}
to obtain 
\begin{equation*}
\mathbf{S}=\lambda\left( 
\begin{array}{cc}
\mathbf{I} & \mathbf{O} \\ 
\mathbf{O} & \mathbf{I}%
\end{array}
\right) +(\lambda+\lambda^{\ast})\left( 
\begin{array}{cc}
-\mathbf{S}_{11} & \mathbf{S}_{12} \\ 
\mathbf{S}_{21} & -\mathbf{S}_{22}%
\end{array}
\right) ,
\end{equation*}
where the matrix elements of $\mathbf{S}$ are given by $\mathbf{S}_{11}=(%
\mathbf{I}+\mathbf{\Psi}_{1}\mathbf{\Psi}_{2}^{-1} \mathbf{\Psi}_{1}^{\ast}%
\mathbf{\Psi}_{2}^{\ast-1})^{-1}$, $\mathbf{S}_{12}=(\mathbf{\Psi }_{2}%
\mathbf{\Psi}_{1}^{-1}+\mathbf{\Psi}_{1}^{\ast}\mathbf{\Psi}_{2}^{\ast
-1})^{-1}$, $\mathbf{S}_{21}=(\mathbf{\Psi}_{2}^{\ast}\mathbf{\Psi}%
_{1}^{\ast-1} +\mathbf{\Psi}_{1}\mathbf{\Psi}_{2}^{-1})^{-1}$, $\mathbf{S}%
_{22}=(\mathbf{I}+\mathbf{\Psi}_{2}\mathbf{\Psi}_{1}^{-1} \mathbf{\Psi}%
_{2}^{\ast}\mathbf{\Psi}_{1}^{\ast-1})^{-1}$.

Finally, the DT for Eq. (\ref{eq1}) follows from relation (\ref{db1}): 
\begin{equation}
\mathbf{Q}_{1}=\mathbf{Q}+2(\lambda+\lambda^{\ast})\mathbf{S}_{12}.
\label{db}
\end{equation}
It should be noted that $\mathbf{S}_{21}^{\ast}=\mathbf{S}_{12}$ implies the
above-mentioned property, $\mathbf{Q}_{1}^{\ast}=\mathbf{Q}_{1}^{\dag}$.
From Eqs. (\ref{db}) and (\ref{lax}), it can be deduced that Eq. (\ref{db})
generates a new solution $\mathbf{Q}_{1}$ for Eq. (\ref{eq1}) once a \textit{%
seed solution} $\mathbf{Q}$ is known. In particular, a one-soliton solution
can be generated if the seed is a trivial (zero) state. Next, taking $%
\mathbf{Q}_{1}$ as the new seed solution, one can derive from Eq. (\ref{db})
the corresponding two-soliton solution. This can be continued as a recursion
procedure generating multi-soliton solutions.

In the following, we will be dealing with solutions to Eq. (\ref{eq1}) under
nonvanishing boundary conditions. The simplest among them is the CW
solution, with constant densities of the three spin components, 
\begin{eqnarray}
\mathbf{Q}_{c} &=&\mathbf{A}_{c}e^{i\varphi _{c}},~\mathbf{A}_{c}\equiv
\left( 
\begin{array}{cc}
\beta _{c} & \alpha _{c} \\ 
\alpha _{c} & -\beta _{c}%
\end{array}%
\right) ,~  \notag \\
\varphi _{c} &\equiv &kx+\left[ 2(\alpha _{c}^{2}+\beta _{c}^{2})-k^{2}%
\right] t,~  \label{CW}
\end{eqnarray}%
where $\alpha _{c}$ and $\beta _{c}$ are real constants, and $k$ is a wave
number. In this solution, the constant densities are equal in components $%
\phi _{\pm 1}$, while the phase difference between the components is $\pi $.

Applying the DT to the CW solution $\mathbf{Q}_{c}$, solving Eqs. (\ref{lax}%
) for this case, and employing relation (\ref{db}), we obtain a new family
of solutions of Eq. (\ref{eq1}) in the form 
\begin{equation}
\mathbf{Q}_{1}=[\mathbf{A}_{c}+4\xi (\mathbf{I}+\mathbf{AA}^{\ast })^{-1}%
\mathbf{A}]e^{i\varphi _{c}},  \label{so}
\end{equation}%
where the following definitions are used: 
\begin{equation}
\mathbf{A}=(\mathbf{\Pi }e^{\theta -i\varphi }+\kappa ^{-1}\mathbf{A}%
_{c})(\kappa ^{-1}\mathbf{A}_{c}\mathbf{\Pi }e^{\theta -i\varphi }+\mathbf{I}%
)^{-1},
\end{equation}%
\begin{align}
\theta & =M_{I}x+[2\xi M_{R}-(k+2\eta )M_{I}]t,  \label{theta} \\
\varphi & =M_{R}x-[2\xi M_{I}+(k+2\eta )M_{R}]t,  \label{varphi}
\end{align}%
\begin{equation}
M=\sqrt{(k+2i\lambda )^{2}+4(\alpha _{c}^{2}+\beta _{c}^{2})}\equiv M_{%
\mathrm{R}}+iM_{\mathrm{I}},  \label{M}
\end{equation}%
$\kappa \equiv \frac{1}{2}(ik-2\lambda +iM)$, $\lambda =\xi +i\eta $ is the
spectral parameter, and $\mathbf{\Pi =}\left( 
\begin{array}{cc}
\beta & \alpha \\ 
\alpha & \gamma%
\end{array}%
\right) $ is an arbitrary complex symmetric matrix. Solution (\ref{so})
reduces back to the CW (\ref{CW}) when $\xi =0$. It is worth noting that the
three-component polar soliton (\ref{polarb3}) considered in the previous
section is not a special example of the solution (\ref{so}), because the
background densities in solution (\ref{polarb3}) have equal phases in
components $\phi _{\pm 1}$.

In particular, with $\mathbf{A}_{c}=\mathbf{O}$ (zero background), Eq. (\ref%
{so}) yields a soliton solution, that can be written as 
\begin{equation}
\mathbf{Q}_{1}=\frac{4\xi\lbrack\mathbf{\Pi}_{1}e^{-\theta_{s}} +(\mathbf{%
\sigma}^{y}\mathbf{\Pi}_{1}^{\ast}\mathbf{\sigma}^{y}) e^{\theta_{s}}\det%
\mathbf{\Pi}_{1}]e^{i\varphi_{s}}}{e^{-2\theta_{s}} +1+e^{2\theta_{s}}|\det%
\mathbf{\Pi}_{1}|^{2}},  \label{so2}
\end{equation}
where $\theta_{s}=2\xi(x-4\eta t)-\theta_{0}$, $\varphi_{s}=2\eta x+4(\xi
^{2}-\eta^{2})t$, and $\theta_{0}$ is an arbitrary real constant which
determines the initial position of the soliton, $\mathbf{\sigma}^{y}$\ is
the Pauli matrix, and $\mathbf{\Pi}_{1}$ is the polarization matrix \cite%
{Ieda}, 
\begin{equation*}
\mathbf{\Pi}_{1}=\left( 2|\alpha|^{2}+|\beta|^{2}+|\gamma|^{2}\right) ^{-1/2}%
\mathbf{\Pi}\equiv\left( 
\begin{array}{cc}
\beta_{1} & \alpha_{1} \\ 
\alpha_{1} & \gamma_{1}%
\end{array}
\right) .
\end{equation*}
Solitons (\ref{so2}) are tantamount to ones found in Ref. \cite{Ieda}, where
they were classified into the above-mentioned distinct types, the FM and
polar ones. Indeed, one can see from expression (\ref{so2}) that, when $\det%
\mathbf{\Pi}_{1}=0$, the soliton (\ref{so2}) represents a FM state in the
spinor model. However, when $\det\mathbf{\Pi}_{1}\neq0$, the solution (\ref%
{so2}) may have two peaks, corresponding to a polar state (the smaller $\det%
\mathbf{\Pi}_{1}$, the larger the distance between the peaks).

We will now produce another particular example of exact solutions (\ref{so}%
), that describes the onset and nonlinear development of the MI
(modulational instability) of CW states. Indeed, under the conditions $%
k=2\eta$ and $\alpha_{c}^{2}+\beta_{c}^{2}>\xi^{2}$, Eq. (\ref{M}) yields $%
M_{\mathrm{I}}=0$, hence $\theta$ is independent of $x$ [see Eq. (\ref{theta}%
)], and the solution features no localization. Instead, it is $x$-periodic,
through the $x$-dependence in $\varphi$, as given by Eq. (\ref{varphi}); in
this connection, we notice that 
\begin{equation}
M_{\mathrm{R}}^{2}=4\left( \alpha_{c}^{2}+\beta_{c}^{2}\right) -4\xi^{2}
\label{MR}
\end{equation}
does not vanish when $M_{\mathrm{I}}=0$. This spatially periodic state may
actually display a particular mode of the nonlinear development of the MI,
in an \emph{exact} form. To analyze this issue in detail, we consider a
special case with $\mathbf{\Pi}=\epsilon\mathbf{E}$, where $\mathbf{E}$ is a
matrix with all elements equal to $1$, and $\epsilon$ is a small amplitude
of the initial perturbation added to a CW to initiate the onset of the MI.
Indeed, the linearization of the initial profile of solution (\ref{so}) in $%
\epsilon$ yields 
\begin{equation}
\mathbf{Q}_{1}(x,0)\approx\lbrack\mathbf{A}_{c}\rho-\epsilon\chi_{1}\mathbf{E%
} \cos(M_{R}x)-\epsilon\chi_{2}\mathbf{\sigma}^{z}\mathbf{\sigma}^{x}\mathbf{%
A}_{c}e^{iM_{R}x}]\exp(ikx),  \label{ini}
\end{equation}
where $\rho\equiv1-\xi(2\xi+iM_{R})/(\alpha_{c}^{2}+\beta_{c}^{2})$ with $%
|\rho|=1$, $\chi_{1}\equiv\xi M_{R}(2i\xi-M_{R})/(\alpha_{c}^{2}+\beta
_{c}^{2})$, $\chi_{2}\equiv\beta_{c}\chi_{1}/(\alpha_{c}^{2}+\beta_{c}^{2})$%
, $M_{R}=2\sqrt{\alpha_{c}^{2}+\beta_{c}^{2}-\xi^{2}}$, and $\mathbf{\sigma }%
^{z,x}$ are the Pauli matrices. Clearly, the first term in Eq. (\ref{ini})
represents the CW, while the second and third ones are small
spatially-modulated perturbations.

Comparing the exact solution (\ref{so}) obtained in this special case with
results of direct numerical simulations of Eqs. (\ref{eq}) with the initial
condition (\ref{ini}) (not shown here), one can verify that the numerical
solution is very close to the analytical one, both displaying the
development of the MI initiated by the small modulational perturbation in
the waveform (\ref{ini}). Figure 8 shows the result in terms of the
densities of the three components, as provided by the exact solution (\ref%
{so}). From this picture we conclude that atoms are periodically
transferred, in the spinor BEC, from the spin state $m_{F}=0$ into ones $%
m_{F}=\pm 1$, and vice versa.

As said above, the exact solution (\ref{so}) to Eqs. (\ref{eq}) is only
valid under the special condition $a=1$ (i.e., $2g_{0}=-g_{2}$). In the
general case ($a\neq 1$), the onset of the MI can be analyzed directly from
equations (\ref{eq}) linearized for small perturbations, as it was actually
done in Ref. \cite{MI}. We will briefly recapitulate the analysis here (in
particular, to compare results with the above exact solution). To this end,
we take the CW solution (\ref{CW}) (with $k=0$, which can be fixed by means
of the Galilean transformation), and consider its perturbed version, 
\begin{eqnarray*}
\widetilde{\mathbf{Q}} &=&(\mathbf{A}_{c}+\mathbf{B})\exp [2i\nu (\alpha
_{c}^{2}+\beta _{c}^{2})t], \\
\mathbf{B} &=&\left( 
\begin{array}{cc}
b_{+1}(x,t) & b_{0}(x,t) \\ 
b_{0}(x,t) & b_{-1}(x,t)%
\end{array}%
\right) ,
\end{eqnarray*}%
where $b_{\pm 1}$ and $b_{0}$ are weak perturbations. Substituting this
expression into Eqs. (\ref{eq}), linearizing them, and assuming the
perturbations in the usual form, 
\begin{eqnarray*}
b_{j}(x,t) &=&b_{j\mathrm{R}}\cos (Kx-\omega t)+ib_{j\mathrm{I}}\sin
(Kx-\omega t), \\
~j &=&+1,0,-1,
\end{eqnarray*}%
where $b_{j\mathrm{R,I}}$ are real amplitudes, the linearized equations give
rise to two branches of the dispersion relation between the wavenumber $K$
and frequency $\omega $ of the perturbation, 
\begin{equation}
\omega ^{2}=K^{2}(K^{2}-4a\alpha _{c}^{2}-4a\beta _{c}^{2});  \label{dis1}
\end{equation}%
\begin{equation}
\omega ^{2}=K^{2}(K^{2}-4\nu \alpha _{c}^{2}-4\nu \beta _{c}^{2}).
\label{dis2}
\end{equation}

The MI sets in when $\omega ^{2}$ given by either expression ceases to be
real and positive. Obviously, this happens if, at least, one condition 
\begin{equation}
K^{2}<4a(\alpha _{c}^{2}+\beta _{c}^{2}),~K^{2}<4\nu (\alpha _{c}^{2}+\beta
_{c}^{2})  \label{KK}
\end{equation}%
holds. Note that these conditions agree with the above exact result, which
corresponds to $a=\nu =1$. Indeed, in that case we may identify $K\equiv M_{%
\mathrm{R}}$, according to Eq. (\ref{ini}), and then Eq. (\ref{MR}) entails
the constraint, $M_{\mathrm{R}}^{2}<4\left( \alpha _{c}^{2}+\beta
_{c}^{2}\right) $, which is identical to conditions (\ref{KK}). Note that
the MI may occur in the case of the repulsive spin-independent interactions
in Eqs. (\ref{eq}), i.e., if $\nu =-1$, provided that the nonlinear-coupling
constant $a$, which accounts for the strength of the spin-dependent
interaction, is positive.

We also explored stability of the soliton solutions (\ref{so2}), of both the
polar and FM types, against finite random variations of the coupling
constant $a$ in time. The issue is relevant, as the exact solutions are only
available for $a=1$, hence it is necessary to understand if the solitons
survive random deviations of $a$ from this special value corresponding to
the integrability (in fact, this way one may test the \emph{structural
stability} of the solitons generated in the exact form by the
inverse-scattering/Darboux transform). Physically, this situation may
correspond to a case when the scattering length, that determines the
nonlinear coefficient $a$ in Eqs. (\ref{eq}), follows random variations of
an external magnetic \cite{FeshbachMagn} or laser \cite{FeshbachOpt} field
which affects the scattering length through the Feshbach resonance. The
evolution of the density profiles in the thus perturbed solitons is
displayed in Fig. 9. This figure and many other runs of the simulations
suggest that the solitons of both the FM and polar types are robust against
random changes of $a$, once they are dynamically stable as exact solutions
to the integrable version of the model. Note also that the FM soliton seems
more robust than its polar counterpart.

\section{Conclusion}

In this work, we have presented new soliton states in the model based on the
coupled NLS equations describing the dynamics of $F=1$ spinor BECs. In the
general (non-integrable) version of the model, several types of elementary
exact soliton solutions were reported, which include one, two, or three
components, and represent polar or ferromagnetic (FM)\ solitons. Their
stability was checked by direct simulations, and, in some cases, in an exact
analytical form, based on linearized equations for small perturbations. The
global stability of the solitons was analyzed by comparing the respective
values of the energy for a fixed norm (number of atoms). As a result, we
have found stable ground-state solitons of the polar and FM types; stable
metastable solitons coexisting with the ground-state ones are possible too,
in certain parameter regions.

In the special case of the integrable spinor-BEC model, we have applied the
Darboux transformation (DT) to derive a family of exact solutions on top of
the CW state. This family includes a solution that explicitly displays full
nonlinear evolution of the modulational instability (MI)\ of the CW state,
seeded by a small spatially periodic perturbation. Finally, by means of
direct simulations, we have established robustness of one-soliton solutions,
of both the FM and polar types, under finite perturbations of the coupling
constant. In fact, the latter result implies the structural stability of
these solitons.

This work was supported by the NSF of China under grants 60490280, 90403034,
90406017, and Overseas Scholar Foundation of Shanxi Province. The work of
B.A.M. was supported, in part, by the Israel Science Foundation through a
grant No. 8006/03.

\bigskip

\textbf{Figure Captions}

Figure 1. Evolution of the single-component polar soliton (\ref{polar1})
with a small random perturbation added, at $t=0$, to the $\phi _{+1}$ and $%
\phi _{-1}$ components. Parameters are $\nu =1$, $a=-0.5$, and $\mu =-1$.

Figure 2. The solid, dashed, and dotted curves show, respectively, the
evolution of the norms $N_{0}$, $N_{\pm }$ of the $\phi _{0}$ and $\phi
_{\pm 1}$ components, and total norm $N$ [see Eqs. (\ref{NN}) and (\ref{N})]
in the two-component polar soliton (\ref{polar2}) perturbed by a small
random perturbation introduced in the $\phi _{0}$ component. Parameters are $%
\nu =1$, $\mu =-1$, and (a) $a=1.5$ for $\phi _{+1}\phi _{-1}^{\ast }<0$;
(b) $a=1.5$ for $\phi _{+1}\phi _{-1}^{\ast }>0$; (c) $a=-1.5$ for $\phi
_{+1}\phi _{-1}^{\ast }<0$; (d) $a=-1.5$ for $\phi _{+1}\phi _{-1}^{\ast }>0$%
.

Figure 3. The evolution of the three-component polar soliton (\ref{polar31})
under the action of a small random perturbation initially added to the $\phi
_{0}$ component. Parameter are $\nu =1$, $\mu =-0.8$, $\epsilon =2/\sqrt{5}$%
, and (a) $a=0.5$; (b) $a=1.5$; (c) $a=-0.5$; (d) $a=-1.5$. The meaning of
the solid, dashed, and dotted curves is the same as in Fig. 2.\ 

Figure 4. The same as in Fig. 3 for the three-component polar soliton (\ref%
{polar32}). Parameters are $\nu =1$, $\mu =-0.8$, $\epsilon =2/\sqrt{5}$,
and (a) $a=0.5$; (b) $a=1.5$; (c) $a=-0.5$; (d) $a=-1.5$. The meaning of the
solid, dashed, and dotted curves is the same as in Figs. 2 and 3.\ 

Figure 5. The evolution of the three-component polar soliton (\ref{polar3a})
with initially added random perturbations. Parameters are $\nu =1$, $\mu
_{+1}=-1.21$, $\mu _{-1}=-0.25$, and$\ $(a,c) $a=0.5$, (b,d) $a=-0.5$. In
cases (a) and (b), the random perturbation was added to the $\phi _{0}$
component, and in cases (c) and (d) -- to components $\phi _{\pm 1}$. The
curves marked by 1, 2, 3, and 4 depict, respectively, the evolution of the
norms $N_{+1}$, $N_{-1}$, $N_{0}$ and $N$, see the definitions in Eqs. (\ref%
{NN}) and (\ref{N}).\ 

Figure 6. The evolution of the two-component polar soliton on the CW
background, (\ref{polarb2}), under the action of initial random
perturbations added to the $\phi _{0}$ and $\phi _{\pm 1}$ components.
Parameters are $\nu =+1,a=-1/2,$ and $\mu =-1$.\ \ \ \ \ 

Figure 7. The same as in Fig. 6, for the finite-background soliton (\ref%
{polarb3}). Parameters are $\nu =a=1,$ and $\mu =-0.36$.\ \ \ \ \ 

Figure 8. The nonlinear development of the modulation instability, as per
the exact solution (\ref{so}) under the conditions $k=2\eta $ and $\alpha
_{c}^{2}+\beta _{c}^{2}>\xi ^{2}$. Parameters are $k=0.6$, $\eta =0.3$, $\xi
=1$, $\alpha =\beta =\gamma =e^{-8}$, $\alpha _{c}=1.2$, and $\beta _{c}=0$%
.\ 

Figure 9.\ The evolution of the density distribution in the soliton obtained
from solution (\ref{so}) by adding a randomly time-dependent perturbation of
the nonlinear-coupling constant $a$, with the perturbation amplitude $\pm
5\% $. Parameters are $\nu =a=1$ (as concerns the unperturbed value of $a$), 
$\xi =1$, $\eta =-0.03$, and $\theta _{0}=0$. (a) The FM soliton, with $%
\alpha _{1}=0.48$, $\beta _{1}$ and $\gamma _{1}$ being determined by
conditions $\alpha _{1}^{2}=\beta _{1}\gamma _{1}$ and $2\left\vert \alpha
_{1}\right\vert ^{2}+\left\vert \beta _{1}\right\vert ^{2}+\left\vert \gamma
_{1}\right\vert ^{2}=1$; (b) the polar soliton, with $\alpha _{1}=0.53$ and $%
\beta _{1}=0.43$, $\gamma _{1}$ being determined by $2\left\vert \alpha
_{1}\right\vert ^{2}+\left\vert \beta _{1}\right\vert ^{2}+\left\vert \gamma
_{1}\right\vert ^{2}=1$. The solid and dotted curves show, respectively, the
exact solutions and the perturbed ones produced by numerical simulations.\ \
\ \ \ \ \ \ \ \ \ \ \ \ \ \ \ \ \ \ \ \ \ \ \ \ 


\begin{thebibliography}{99}
\bibitem{Burger} S. Burger, K. Bongs, S. Dettmer, W. Ertmer, K. Sengstock,
A. Sanpera, G. V. Shlyapnikov, and M. Lewenstein,\textit{\ }Phys. Rev. Lett. 
\textbf{83}, 5198 (1999); J. Denschlag, J. E. Simsarian, D. L. Feder,
Charles W. Clark, L. A. Collins, J. Cubizolles, L. Deng, E. W. Hagley, K.
Helmerson, W. P. Reinhardt, S. L. Rolston, B. I. Schneider, and W. D.
Phillips, Science \textbf{287}, 97 (2000).

\bibitem{bright} K. E. Strecker, G. B. Partridge, A. G. Truscott, and R. G.
Hulet, Nature (London) \textbf{417}, 150 (2002); L. Khaykovich, F. Schreck,
G. Ferrari, T. Bourdel, J. Cubizolles, L. D. Carr, Y. Castin, and C.
Salomon, Science \textbf{296}, 1290 (2002).

\bibitem{gap} B.~Eiermann, Th. Anker, M. Albiez, M. Taglieber, P. Treutlein,
K. -P. Marzlin, and M. K. Oberthaler, Phys. Rev. Lett. \textbf{92}, 230401
(2004).

\bibitem{Busch} Th. Busch and J. R. Anglin, Phys. Rev. Lett. \textbf{84},
2298 (2000); L. Salasnich, Phys. Rev. A\textbf{70}, 053617 (2004).

\bibitem{Rosenbusch} P. Rosenbusch, V. Bretin, and J. Dalibard, Phys. Rev.
Lett. \textbf{89}, 200403 (2002).

\bibitem{Liu} W. M. Liu, B. Wu, and Q. Niu, Phys. Rev. Lett. \textbf{84},
2294 (2000).

\bibitem{DW} M. Trippenbach \textit{et al}, 
J. Phys. B \textbf{33}, 4017 (2000); B. A. Malomed, H. E. Nistazakis,, D. J.
Frantzeskakis, and P. G. Kevrekidis, Phys. Rev. A \textbf{70}, 043616
(2004); M. I. Merhasin, B. A. Malomed, and R. Driben, J. Phys. B \textbf{38}%
, 877 (2005).

\bibitem{Salasnich} L. Salasnich, A. Parola, and L. Reatto, Phys. Rev. Lett. 
\textbf{91}, 080405 (2003); K. Kasamatsu, and M. Tsubota, Phys. Rev. Lett. 
\textbf{93}, 100402 (2004).

\bibitem{Carr} L. D. Carr, and J. Brand, Phys. Rev. Lett. \textbf{92},
040401 (2004).

\bibitem{Miesner} H. J. Miesner, D. M. Stamper-Kurn, J. Stenger, S. Inouye,
A. P. Chikkatur, and W. Ketterle, Phys. Rev. Lett. \textbf{82}, 2228 (1999).

\bibitem{Ohmi} M. Ohmi, and K. Machida, J. Phys. Soc. Jpn. \textbf{67}, 1822
(1998).

\bibitem{Stenger} J. Stenger, S. Inouye, D. M. Stamper-Kurn, H. -J. Miesner,
A. P. Chikkatur, and W. Ketterle, Nature (London) \textbf{396}, 345 (1998).

\bibitem{Stamper} D. M. Stamper-Kurn, M. R. Andrews, A. P. Chikkatur, S.
Inouye, H.-J. Miesner, J. Stenger, and W. Ketterle, Phys. Rev. Lett. \textbf{%
80}, 2027 (1998).

\bibitem{Ho} T. L. Ho, Phys. Rev. Lett. \textbf{81}, 742 (1998).

\bibitem{Pu} H. Pu, C. K. Law, S. Raghavan, J. H. Eberly, and N. P. Bigelow,
Phys. Rev. A\textbf{60}, 1463 (1999).

\bibitem{Ho1} C. V. Ciobanu, S.-K. Yip, and Tin-Lun Ho, Phys. Rev. A\textbf{%
61}, 033607 (2000).

\bibitem{Koashi} M. Koashi, and M. Ueda, Phys. Rev. Lett. \textbf{84}, 1066
(2000).

\bibitem{MI} N. P. Robins, W. Zhang, E. A. Ostrovskaya, and Y. S. Kivshar,
Phys. Rev. A \textbf{64}, 021601(R) (2001).

\bibitem{Ueda} M. Ueda, and M. Koashi, Phys. Rev. A\textbf{65}, 063602
(2002).

\bibitem{Schmaljohann} H. Schmaljohann, M. Erhard, J. Kronjager, M. Kottke,
S. van Staa, L. Cacciapuoti, J. J. Arlt, K. Bongs, and K. Sengstock, Phys.
Rev. Lett. \textbf{92}, 040402 (2004).

\bibitem{Xie} Z. W. Xie, W. P. Zhang, S. T. Chui, and W. M. Liu, Phys. Rev. A%
\textbf{69}, 053609 (2004).

\bibitem{Ieda} J. Ieda, T. Miyakawa, and M. Wadati, Phys. Rev. Lett. \textbf{%
93}, 194102 (2004), and references therein.

\bibitem{Agrawal} G. P. Agrawal. \textit{Nonlinear Fiber Optics} (Academic
Press: San Diego, 1995).

\bibitem{Barash} I. V. Barashenkov, M. M. Bogdan, and V. I. Korobov,
Europhys. Lett. \textbf{15}, 113 (1991).\newline

\bibitem{HH} J. -C. van der Meer, Nonlinearity \textbf{3}, 1041 (1990); P.
D. Woods, and A. R. Champneys, 
Physica D 129, 147 
(1999); B. A. Malomed and R. S. Tasgal, Phys. Rev. E \textbf{49}, 5787
(1994); I. V. Barashenkov, D. E. Pelinovsky, and E. V. Zemlyanaya, Phys.
Rev. Lett. \textbf{80}, 5117 (1998); D. Mihalache, D. Mazilu, and L. Torner,
Phys. Rev. Lett. \textbf{81}, 4353 (1998); D. Mihalache, D. Mazilu, and
L.-C. Crasovan, Phys. Rev. E \textbf{60}, 7504 (1999).

\bibitem{Manakov} S. V. Manakov, Sov. Phys. JETP \textbf{38}, 248 (1974).

\bibitem{VK} M. G. Vakhitov and A. A. Kolokolov, Izv. Vuz. Radiofiz. \textbf{%
16}, 1020 (1973) [in Russian; English translation: Sov. J. Radiophys.
Quantum Electr. \textbf{16}, 783 (1973)]; L. Berg{\'{e}}, Phys. Rep. \textbf{%
303}, 260 (1998).

\bibitem{Tsuchida} T. Tsuchida, and M. Wadati, J. Phys. Soc. Jpn. \textbf{67}%
, 1175 (1998).

\bibitem{FeshbachMagn} S.~Inouye, M. R. Andrews, J. Stenger, H.-J. Miesner,
D. M. Stamper-Kurn, and W. Ketterle, Nature \textbf{392}, 151 (1998);
E.~A.~Donley, N. R. Claussen, S. L. Cornish, J. L. Roberts, E. A. Cornell,
and C. E. Wieman, Nature \textbf{412}, 295 (2001).

\bibitem{FeshbachOpt} P. O. Fedichev, Yu. Kagan, G. V. Shlyapnikov, and J.
T. M. Walraven, Phys. Rev. Lett. \textbf{77}, 2913 (1996); M. Theis, G.
Thalhammer, K. Winkler, M. Hellwig, G. Ruff, R. Grimm, and J. H. Denschlag,
Phys. Rev. Lett. \textbf{93}, 123001 (2004).
\end{thebibliography}
\end{document}